# Maximum work for Carnot-like heat engines with infinite heat source

Rui Long and Wei Liu*

School of Energy and Power Engineering, Huazhong University of Science and Technology,

Wuhan 430074, China

Corresponding author: Wei Liu (w_liu@hust.edu.cn)

An analysis of efficiency and its bounds at maximum work output for Carnot-like heat engines is conducted and the heat transfer processes are described by Newton's law of cooling with time-dependent heat conductance. The upper bound of the efficiency is found to be the CA efficiency, and is independent of the time duration completing each process and the time-dependent conductance. We prove that even the working medium exchanges heat sufficiently with the heat reservoirs, the work which could be extracted is finite and limited by $W^{max+} = cmT_h \eta_{CA}^2$. The optimal temperature profiles in the heat exchanging processes are also analyzed. When the dimensionless contact times satisfy certain relations, the endoreversible model is recovered.

As we all know, the Carnot efficiency, $\eta_C = 1 - T_c / T_h$, defines the upper bound of efficiency of all the heat engines operating between two heat reservoirs at temperatures $T_h$ and $T_c$, $T_h > T_c$ [1]. Any real-life heat engines should operate at the efficiency lower than $\eta_C$. However, in the Carnot heat engines all the processes are qusi-static and the time for completing that cycle is infinitely long which leads to zero power output. As to actual heat engines, its time duaraion is finite, which shows a significant deviative from the ideal Carnot ones. The ideal carnot cycle must be speeded up to meet the actual demand. Taking only into consideration the irreversibility caused by heat transfer between the heat reservoirs and working substance during the isothermal processes, Curzon-Ahlborn[2] proposed the concept of endoreversible Carnot heat engine, and deduced its efficiency at maximum power output. That is the groundbreaking CA efficiency $\eta_{CA} = 1 - \sqrt{T_c / T_h}$. It opens the era of finite time thermodynamics. In addition, Otto and J-B cycles operating at the maximum work (MW) output also have the efficiencies as $\eta_{CA}$ and Diesel and Atkinson cycles have MW efficiencies very close to $\eta_{CA}$ [3, 4].

For actual heat engines, in the heat aborsbing process the temperature of the working medium should be lower than that of the hot reservoir and in the heat releasing process the temperature of the working medium should be higher than that of the cold reservoir to estabilish the heat exchanging processes. The heat exchanging processes are no longer isothermal. To discribe the heat exchanging between the working meium and the heat reserviors, different heat transfer laws have been systematically studied [5-7]. The geneural heat transfer law is depicted as $\dot{Q} = k(T_s^n - T^n)$, where $k$ is the heat conductance; $T_s$ and $T$ are the temperatures of the heat reservoir and the working medium, respectively. When $n=1$, it returns to the linear transfer law[8, 9], and the Stefan-boltzmann radiation law if $n=4$. Therefore it is of universality. In addition the Dulong-Petit nonlinear heat transfer law is also investigated [10]. However all the heat transfer laws studied before assume the heat conductance stays constant during the heat exchanging process. For actual heat transfer process, as the development of the heat transfer, the temperature difference between the working medium and the heat reservoir is decreasing. Therefore the heat conductance should decrease as well. The heat conductance should not be kept constant, rather be a time-dependent decreasing variable. In this paper, we assume the heat transfer law between the heat reservoir and the working medium confirms to the linear law with time-dependent heat conductance. The efficiency at the maximum work output is deduced. And the maximum work which can be extracted is also proposed.

As to heat engines, a certain amount of heat $Q_h$ is absorbed from the hot reservoir ($T_h$), and some of which $Q_c$, is evacuated to the cold reservoir ($T_c$) at the end of a cycle. The heat transfer law between the heat source and the working medium is assumed to conform to Newton's law of cooling:

$$\frac{dQ}{dt} = cm\frac{dT}{dt} = k(T_s - T) \tag{1}$$

where $c$ is the heat capacity, $m$ is the working substance mass, $T$ is the working substance temperature, $T_s$ is heat source temperature, $k$ is the heat conductance. In this paper, we assume the heat conductance between the working medium and the hot and cold reservoirs conform to the following power-law relations: $k_h = a(\frac{T_h - T}{T_h - T_{h0}})^n$, $k_c = b(\frac{T - T_c}{T_{c0} - T_c})^n$, $n > 0$, where the subscripts ($h$ and $c$) indicate the hot and cold reservoirs; $a$, $b$, $T_{h0}$ and $T_{c0}$ are the initial heat conductance and initial temperatures of the working substance at the beginning of the heating and cooling processes, respectively. According to Eq.(1), the working substance temperature in the heat absorbing process is a function of time $t$:

$$T_{hw}(t) = T_h + (T_{h0} - T_h)(n\tau_h / \Psi_h + 1)^{-1/n} \qquad (2)$$

where $\Psi_h = cm/a$, which reflects the temperature increase degree of the working medium in the time absorbing process and has the dimension of time. The time duration is denoted as $\tau_h$, the heat absorbed from the hot reservoir can be calculated as:

$$Q_h = \int_0^{\tau_h} k_h(T_h - T_{hw}) = cm(T_h - T_{h0})[1 - (n\tau_h / \Psi_h + 1)^{-1/n}] \qquad (3)$$

The relative entropy change of the working substance in the heat absorbing process is given by:

$$\Delta s_h = \int_0^{\tau_h} \frac{dQ_h}{T} = cm \ln \frac{T_h + (T_{h0} - T_h)(n\tau_h / \Psi_h + 1)^{-1/n}}{T_{h0}} \qquad (4)$$

Similarly, the temperature of the working medium, the heat evacuated to the cold reservoir and the entropy change during the heat releasing process are given by

$$T_{cw}(t) = T_c - (T_c - T_{c0})(nt / \Psi_c + 1)^{-1/n} \qquad (5)$$

$$Q_c = \int_0^{\tau_c} k_c(T_c - T_{cw}) = cm(T_c - T_{c0})[1 - (n\tau_c / \Psi_c + 1)^{-1/n}] \qquad (6)$$

$$\Delta s_c = \int_0^{\tau_c} \frac{dQ_c}{T} = -cm \ln \frac{T_c - (T_c - T_{c0})(n\tau_c / \Psi_c + 1)^{-1/n}}{T_{c0}} \qquad (7)$$

where $\Psi_c = cm/b$, which reflects the temperature increase degree of the working medium in the time releasing process and has the dimension of time. $\tau_c$ is the time duration of that process. In this paper, we assume that the compressing and expanding processes are isentropic and the times for completing those processes are zero. After a cycle, the working substance return to its initial state, and the total entropy change of the working substance should be zero *i.e.* $\Delta s_h + \Delta s_c = 0$. Then we have

$$\frac{T_c - (T_c - T_{c0})(n\tau_c / \Psi_c + 1)^{-1/n}}{T_{c0}} \frac{T_h + (T_{h0} - T_h)(n\tau_h / \Psi_h + 1)^{-1/n}}{T_{h0}} = 1 \qquad (8)$$

The work extracted during the cycle is $W = Q_h + Q_c$, and the efficiency is $\eta = 1 + Q_c / Q_h$. Combining Eq.(8) and maximizing $W$ with respect to $T_{c0}$, we have

$$T_{c0}{}^{opt} = \frac{1-(\frac{n\tau_h}{\Psi_h}+1)^{-1/n}+\sqrt{1-\eta_C}(\frac{n\tau_h}{\Psi_h}+1)^{-1/n}[1-(\frac{n\tau_c}{\Psi_c}+1)^{-1/n}]}{\sqrt{1-\eta_C}\{1-[(\frac{n\tau_h}{\Psi_h}+1)(\frac{n\tau_c}{\Psi_c}+1)]^{-1/n}\}} T_c \tag{9}$$

Substituting Eq.(9) into Eq.(8), we can also obtain the optimal initial temperature of the working medium in the heat releasing process.

$$T_{h0}{}^{opt} = \frac{\sqrt{1-\eta_C}[1-(\frac{n\tau_c}{\Psi_c}+1)^{-1/n}]+(\frac{n\tau_c}{\Psi_c}+1)^{-1/n}[1-(\frac{n\tau_h}{\Psi_h}+1)^{-1/n}]}{1-(\frac{n\tau_c}{\Psi_c}+1)^{-1/n}+(\frac{n\tau_c}{\Psi_c}+1)^{-1/n}[1-(\frac{n\tau_h}{\Psi_h}+1)^{-1/n}]} T_h \tag{10}$$

From the above equations, we can obtain the efficiency at MW $\eta_m = 1-\sqrt{1-\eta_C} \equiv \eta_{CA}$. It is independent of the time duration in either process, and is the CA efficiency, which has been obtained though and JB cycles at MW [3, 4, 11]. Although the upper bounds of the efficiency are the same, our model is more universal and general. our model can decribe the J-B and Otto cycles if We let $c = c_p$, $c = c_v$, respectively. What is more, this model does not specify the thermodynamic pathes in the heat exchanging process. Therefore it can desicribe any arbitraty thermodynamic path of the heat exchanging process and is more practical and realistic than the traditional ones.

We can get the temperature profiles in the heat absorbing and releasing process by substituting Eqs.(10) into Eq.(2) and substituting Eq. (9) into Eq.(5). The work output can be rewritten as

$$\begin{cases} W^{max} = cmT_h\eta_{CA}[1-(\frac{n\tau_c}{\Psi_c}+1)^{-1/n}][1-(\frac{n\tau_h}{\Psi_h}+1)^{-1/n}] \\ (\frac{1}{1-(\frac{n\tau_c}{\Psi_c}+1)^{-1/n}+(\frac{n\tau_c}{\Psi_c}+1)^{-1/n}[1-(\frac{n\tau_h}{\Psi_h}+1)^{-1/n}]} - \frac{\sqrt{1-\eta_C}}{1-[(\frac{n\tau_h}{\Psi_h}+1)(\frac{n\tau_c}{\Psi_c}+1)]^{-1/n}}) \end{cases} \tag{11}$$

Eq.(11) achieves its maximum value $W^{max+} = cmT_h\eta_{CA}{}^2$ when $\tau_h/\Psi_h \to \infty$ and $\tau_c/\Psi_c \to \infty$. It indicates that even the working medium exchanges heat sufficiently with the heat reservoirs, the work which could be extracted is finite and limited by $W^{max+}$ which is determined by the heat capacity and the temperatures of hot and cold reservoirs. While in ideal Carnot heat engines, there is no limitation of work output, other than the efficiency. According to Eq.(11), we have the power output at the maximum work $P = W^{max}/(\tau_h+\tau_c)$. Although when $\tau_h/\Psi_h \to \infty$ and

$\tau_c / \Psi_c \to \infty$, $W^{max}$ achieves it maximum value, but the efficiency stays unchanged and is still the CA efficiency. However its power output is zero.

We define $\tau / \Psi$ as the dimensionless contact time, which reflects the equilibrium degree of the temperature between the working medium and heat reservoirs. According to Eqs.(9) and (10), when $\tau_h / \Psi_h \to 0$ and $\tau_c / \Psi_c \neq 0$, we have $T_{c0}{}^{opt} = T_c$ and $T_{h0}{}^{opt} = \sqrt{1-\eta_C} T_h$. For $\tau_h / \Psi_h \to 0$, the heat absorbing processes are short enough so that the final temperature of the working substance is almost equal to its initial temperature. Furthermore $T_{c0} \geq T_{cw}(t) \geq T_c$ to establish the heat releasing process. Then we have $T_{cw}(t)^{opt} = T_c$ and $T_{hw}(t)^{opt} = T_{h0}{}^{opt} = \sqrt{1-\eta_C} T_h$. It means the heat exchanging processes are isothermal.

When $\tau_c / \Psi_c \to 0$ and $\tau_h / \Psi_h \neq 0$, we have $T_{c0}{}^{opt} = T_c / \sqrt{1-\eta_C}$ and $T_{h0}{}^{opt} = T_h$. For $\tau_c / \Psi_c \to 0$, the heat releasing process is short enough so that the final temperature of the working substance is almost equal to its initial temperature. Furthermore $T_h \geq T_{hw}(t) \geq T_{h0}$ to establish the heat exchanging process. Then we have $T_{hw}(t)^{opt} = T_h$ and $T_{cw}(t)^{opt} = T_{c0}{}^{opt} = T_c / \sqrt{1-\eta_C}$, It means that the temperatures of the working medium also keep constant in the heat exchanging processes. But the temperature profiles are not like those obtained under the situation: $\tau_h / \Psi_h \to 0$ and $\tau_c / \Psi_c \neq 0$.

For situations where $\tau_c / \Psi_c \to 0$ and $\tau_h / \Psi_h \to 0$, the final temperatures are the same as the initial ones. Therefore $T_{hw}(t)^{opt} = T_{h0}{}^{opt}$, $T_{cw}(t)^{opt} = T_{c0}{}^{opt}$. The heat conductance are also kept constant and are the initial ones, respectively. However the limits of Eqs.(9) and (10) do not exit. We assume the time durations fulfill the relation $\tau_h / \tau_c = \sqrt{b/a}$ which is obtained through the endoreversible model under the maximum power output [2]. Applying the limit $\tau / \Psi \to 0$, we also deduce the same temperature profiles as those in Ref. [2]. Meanwhile $P$ also achieves its maximum value. Furthermore the corresponding efficiency is also $\eta_{CA}$. That is to say, the endoreversible model is recovered under these situations.

In addition, when $\tau_h / \Psi_h \to \infty$, $\tau_c / \Psi_c \neq 0$, we have $T_{c0}^{opt} = T_c / \sqrt{1-\eta_C}$. The heat absorbing process is long enough so that the final temperature of the working substance is almost equal to that of the heat reservoir, *i.e.* $T_{hw}(\tau_h) = T_h$. And its initial temperature is

$$T_{h0}^{opt} = \{\sqrt{1-\eta_C}[1-(\frac{n\tau_c}{\Psi_c}+1)^{-1/n}] + (\frac{n\tau_c}{\Psi_c}+1)^{-1/n}\} T_h \tag{12}$$

It depends on the contact time of the cold reservoir. For situations where $\tau_c / \Psi_c \to \infty$, $\tau_h / \Psi_h \neq 0$, we have $T_{h0}^{opt} = \sqrt{1-\eta_C} T_h$. The heat releasing process is long enough so that the final temperature of the working substance is almost equal to that of the heat reservoir, i.e. $T_{cw}(\tau_c) = T_c$. And its initial temperature is

$$T_{c0}^{opt} = \frac{(\sqrt{1-\eta_C}-1)(\frac{n\tau_h}{\Psi_h}+1)^{-1/n}+1}{\sqrt{1-\eta_C}} T_c \tag{13}$$

It depends on the contact time of the hot reservoir. Furthermore, when $\tau_h / \Psi_h \to \infty$ and $\tau_c / \Psi_c \to \infty$, we could obtain the initial temperatures, $T_{c0}^{opt} = T_c / \sqrt{1-\eta_C}$, $T_{h0}^{opt} = \sqrt{1-\eta_C} T_h$, and the final temperatures $T_{cw}(\tau_c) = T_c$, $T_{hw}(\tau_h) = T_h$ for the heat absorbing and releasing processes, respectively.

Representative cases have been studied to investigate the impact of power index *n* on the optimal heat engine cycles as depicted in Fig. 1. The optimal temperature profile of the working medium in the heat absorbing process is nearly a linear function of *S* and moves upward with increasing *n*. As we all know, temperature profile should be concave, and to the most extend can be linear in the *T-S* diagram. Therefore, the heat absorbing process approaches its maximum ability in our optimal conditions. Under the same dimensionless contact time, the lower *n* leads to a lager heat conductance, thus more work could be extracted. The temperature differences between the heat reservoirs and working medium in both heat exchanging processes are very significant, from which the irreversible entropy generation stems. In addition, by considering various *n*, we could have a further insight into other cycles whose temperature profiles during the heat exchanging processes are not constant such as J-B, Otto cycles and any other cycles with isentropic compression and expansion processes.

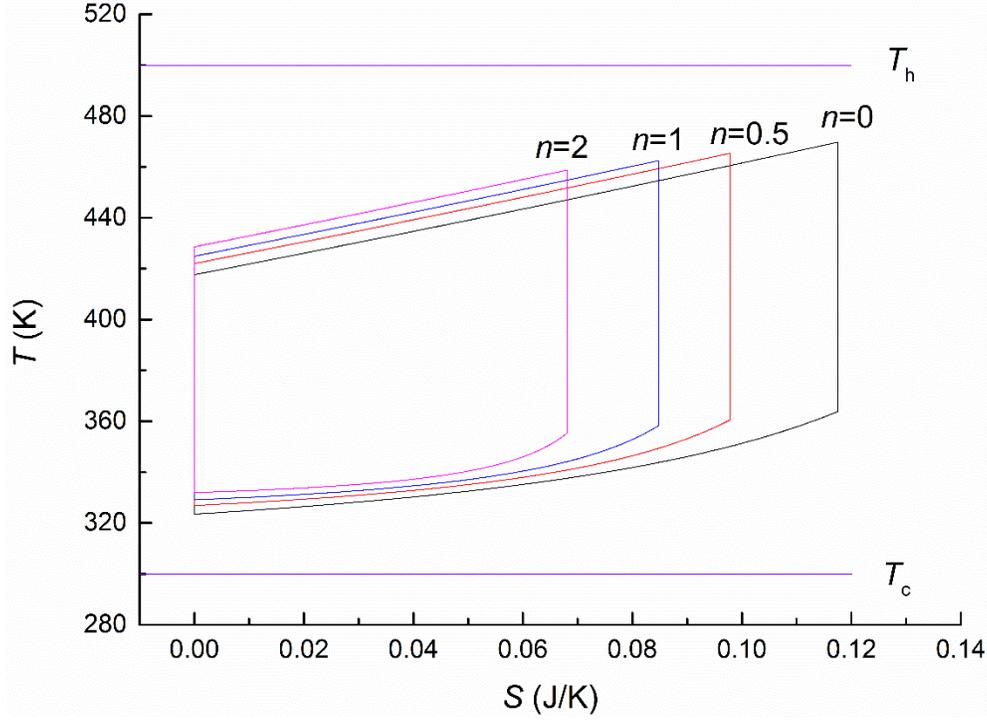

FIG. 1 The T-S diagrams of different optimal heat engine cycles under different $n$ ($n = 0, 0.5, 1, 2$), where $T_h = 500\text{K}$, $T_c = 300\text{K}$, $\tau_h / \Psi_h = \tau_c / \Psi_c = 1$ and $cm = 1\text{J/K}$

In conclusion, we have conducted as analysis of efficiency and its bounds at maximum work output for Carnot-like heat engines whose heat transfer processes are described by Newton's law of cooling. But for generality, the heat conductance are no longer treated as constants, rather as time-dependent decreasing variables. The upper bound of the efficiency is found to be the CA efficiency, and is independent of the time duration completing either process or the time-dependent conductance. Furthermore even the working medium exchanges heat sufficiently with the heat reservoirs, the work which could be extracted is finite and limited by $W^{max+} = cmT_h \eta_{CA}^2$. The optimal temperature profiles in the heat exchanging processes are analyzed under different dimensionless contact time limits. When the dimensionless contact times satisfy certain relations, the endoreversible model is recovered. Furthermore representative cases have been studied to investigate the effect of $n$ on the optimal temperature profiles. The results in present paper could offer us a further insight into any heat engine cycles with isentropic compression and expansion processes. This might be of great guidance for designing or operating actual heat engines.

The authors are grateful for the financial support from the National Natural Science Foundation of China (No.51036003) and the National Key Basic Research Development Program of China (No.2013CB228302).

---